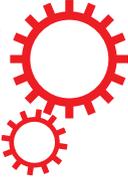

# OPEN

# A novel EM concentrator with open-concentrator region based on multi-folded transformation optics

Hamza Ahmad Madni[1,2], Khurram Hussain[1], Wei Xiang Jiang[1], Shuo Liu[1], Asad Aziz[1], Shahid Iqbal[1], Athar Mahboob[3] & Tie Jun Cui[1]

Conventional concentrators with inhomogeneous coating materials that fully enclose the destined region pose great challenges for fabrication. In this paper, we propose to design an EM concentrator with homogeneous materials. Distinguished from conventional ones, the elaborately designed EM concentrator features a concentrator region that is open to the outer-world, which is achieved with multi-folded transformation optics method by compressing and folding the coating materials to create window(s). Based on this concept, we also investigate open-rotator and open rotational-concentrator devices, which could simultaneously rotate and store the EM waves in the central destined region. Due to the open nature of our proposed designs, we believe they will find potential applications in remote controlling with impressive new functionalities.

The mathematical tool of transformation optics (TO)[1–6] has proved that any object can be made invisible for electromagnetic (EM) waves if the object is fully coated with properly designed inhomogeneous anisotropic[1] or isotropic inhomogeneous[2] materials. Besides the invisibility cloak[7–16], EM field-rotator[17–20] and field-concentrator[21–31] have attained remarkable attention that can rotate the internal propagation direction of EM waves, and increases the EM energy in the inner region, respectively, while the external field remains undisturbed. Therefore, the foregoing development in TO-based devices[1–31] endows materials with very complex properties[5,8,18] and on the other hand, the demands of manipulating the EM waves in the sense of remote control is rapidly increasing in modern world that require compact open-devices with good exposure, easy to upgradeable, and importance of inner-outer world[32–34].

Start from concentrator[21], Fabry Perot resonances were used to design and fabricate the concentrators[35] that achieved progressive attraction in novel energy devices. In addition, some unidirectional and angle-dependent based concentrators[29] have been investigated that work efficiently only in some specific incident wave-angles. Beyond this, rotators[20] are designed for the polarization and have wide applications in the antennas and wave-guiding that provides a reason of linearly polarized into a circularly polarized nature[18]. By taking the combine effects of concentrators and rotators, this phenomenon can be made in practice for Smart Grid applications[36], large area wide band imaging[37], commonly in laser LED-pumped devices[38] and also in micro solar cells arrays[39]. Specifically, parabolic concentrators provides practical impacts in radio broadcasting & motion picture recording[40], while polarization based rotators manipulates effects in some add-drop filter systems[41]. Meanwhile, the TO-based EM concentrators and rotators yield straight path in the manufacturing and applications of metamaterials[18].

Despite all these facts EM concentrators and rotators possess few discrepancies by representing some known reasons. As metamaterials provides very low practical outputs due to their low characteristics and narrow bandwidths[5,8,18]. Previous work concluded that in the designing of the TO based devices, the material parameters were set in inhomogeneous coordinate transformation which is hard to predict the realization phenomenon and difficult for fabrication[42,43]. Beside this, early EM concentrators and rotators show complete invisibility of the destined (concentration and/or rotator) region because of the designing method of these conventional devices that involves in the construction of enclosures the destined region by coating with properly designed materials[17–31].

[1]State Key Laboratory of Millimeter Waves, Department of Radio Engineering, Southeast University, Nanjing, 210096, China. [2]Department of Computer Engineering, Khwaja Fareed University of Engineering & Information Technology, Rahim Yar Khan, 64200, Pakistan. [3]Department of Electrical Engineering, Khwaja Fareed University of Engineering & Information Technology, Rahim Yar Khan, 64200, Pakistan. Correspondence and requests for materials should be addressed to H.A.M. (email: 101101770@seu.edu.cn) or T.J.C. (email: tjcui@seu.edu.cn)

However, in such conventional recipes, even a pinhole gap in the coating material will lead a deterioration of desired functionality. To solve these issues, we have the necessity to design such EM concentrators and rotators, which provide clear path for the better outcomes[18] when the destined region is not fully coated with any EM medium constructed by homogeneous materials that are easy to fabricate.

Now the question raised that whether we can manipulate the EM waves when the air-gap in the homogenous coating materials exists. As an answer, in this paper, we propose an easy way to design "open-coating" devices by utilizing the linear homogenous transformation method and this interesting phenomenon is associated with concentrators, field rotators, and rotational-concentrator. Our proposed work is not only able to rotate but also store the EM waves simultaneously and most importantly, the destined region is open to outer world.

Unlike inhomogeneous transformation, here, we divide the imaginary space into several triangular shaped segments. Our proposed idea is based on the compression of coating materials and then map it to the transformed medium, which leads to creation of window(s) in front of the destined region and the constitutive parameters of the proposed device can be obtained by applying multi-folded TO[33,34]. To understand and validate the proposed concept, we will consider three different examples. In the first example, we introduce a different scheme to design concentrator with homogenous coating-materials that contains air-gaps in front of the concentrator region but the functionality remains unchanged. In this case, the concentrating region will become open to outer-world and can be easily used for matter exchange without scratching the whole coating-material. Based on this, we extend our idea to construct open-coating rotators to artificially change the propagating direction of EM waves when the rotation-region is open to outer world. Finally yet importantly, we proposed an open-coating bi-functional device that can rotate and store the EM waves simultaneously. The proposed open-coating concept is devoted to remote controlling applications of EM waves and in the other fields of engineering and in each example; the proposed open-coated device is placed at a certain distance from the destined region. In the following, full wave finite element method is used to demonstrate and validate the expected behavior of our proposed concepts.

## Results and Discussions

Conventional concentrators[13,21] are based on the radial mapping that store the EM energy in the smaller concentrator region. Meanwhile the concentrator device itself is also the example of invisibility as there is no scattering for the outside observers. Therefore, the expanding technique is applied in ref.[21] and the folding method is used in ref.[13] to design complementary media based concentrator. Thus, in previous works, the properly designed coating material is used to fully enclose the destined region and that coating material is composed by anisotropic and inhomogeneous materials that are difficult for practical implementations.

Distinguishing from the aforesaid methods, in our approach, we divide the virtual space of $N$ segments into $2N$ triangles and then apply two steps to obtain the desired goal. Figure 1 depicts the schematic diagram of the proposed open-coating concentrator device in which a square shaped virtual space of total four segments are further divided into eight different triangular regions labeled as 1, 2, 3 … 8. In the first step, the region 1 of $\Delta ABC$ and the region 2 of $\Delta CDB$ in the virtual space $(x, y, z)$ is folded into region $\Delta A'B'C$ labeled as $1'$ and $\Delta CDB'$ labeled as $2'$ in the physical space $(x', y', z')$, respectively. Similarly, the other regions in virtual space labeled by 2, 3 … 8 are accordingly folded into colored regions $2', 3'… 8'$ to design the concentrator. It can be seen in Fig. 1, that the air-gap came to exist between the concentrator-region and concentration-coating. Thus, by taking advantage of this facility, in the second step, a compression and folding method is applied by compressing the bigger dashed lines into the compressed region I. For example, here, we applied compression technique to regions $7', 8'$ and the whole segment is compressed into regions as $7''$ and $8''$ respectively, so that the coating materials turn into open that do not require to enclose the concentration-region.

In the second step of Fig. 1, consequently, we compressed the whole black dashed lined space into the region I (represented by $(x_1, y_1, z_1)$) and region II (denoted by $(x_2, y_2, z_2)$). Whereas, the material parameters of region II are obtained from the first step of Fig. 1. Thereafter, to overcome the discontinuity occur due to the compression, the compressed region is further coated with the folded regions such as region III (denoted by $(x_3, y_3, z_3)$), region IV (denoted by $(x_4, y_4, z_4)$), and region V (denoted by $(x_5, y_5, z_5)$). For detail, taking the second quadrant as an example of region III, the $\Delta PQA'$ is folded into $\Delta PQR$ which can be seen in the purple color of zoom-in view. Similarly, region IV and region V in the first quadrant are obtained by folding the space of $QtA'C$ into the light green colored region of $QtRS$ and $\Delta P_1tC$ into the golden colored region of $\Delta P_1tS$. Due to the symmetric structure, other parts of these complementary regions in third and fourth quadrants can be achieved by rotating all of the corresponding tensors by $\pi$. The materials used in folded regions are known as complementary materials that are also used in lens designs to compensate the space[44]. $x_n, y_n,$ and $z_n$ indicate the coordinate system of each region, where $n = 1, 2, 3, 4$ and $5$.

In the following, full wave simulations of finite element method (COMSOL) are performed in 2D by adopting the transverse electric (TE) mode at the frequency of $3GHz$. The geometric structure parameter for the both steps in SI units are as follows: the total width of the squares $a = 0.17, b = 0.21, c = 0.25$ and the coordinates of $A(-0.085, 0.085), B(0.085, 0.085), C(-0.105, 0.105). D(0.105, 0.105), A'(-0.125, 0.125), B'(0.125, 0.125), P(-0.21, 0), P_1(-0.09, 0), Q(-0.125, 0.1), R(-0.125, 0.0875), t(-0.105, 0.084),$ and $S(-0.105, 0.0735)$. The obtained material parameters for each region can be seen in method summary.

In Fig. 2, we consider to examine the working performance of proposed open-coating concentrator with the proposed close-coating concentrator under the illumination of plane waves with the two different incident angles of 0° and 90°, respectively. The electric field $(E_z)$ distributions in all cases (Fig. 2(a–c)) indicate that the electric field's amplitude in the concentrator region is equivalent to that in free space near the outer boundary of concentrator. In Fig. 2a, the working functionality of the designed closed-coating concentrator is similar to that in all the previous reported concentrators[21]. The beauty of proposed concentrator is that the $z-$ direction space is unchanged and concentrator region compresses the space in $x - y$ plane only. Therefore, the total energy flow is towards the concentration region and that region is fully covered with the properly designed coated materials, as shown in Fig. 2a. Another case is observed in Fig. 2b, to design a concentrator with open-coating materials and by comparing the field patterns of Fig. 2a and Fig. 2b, it is evident that the concentration region can be made open for the outside observers while the overall performance remains unchanged. Furthermore, to prove the proposed device have no limitations of incident waves' direction, the incident waves have applied at 90° angle, as seen in Fig. 2c. The field patterns of Fig. 2c is identical to Fig. 2b and this result is the bonus point to validate the proposed concept of this paper. Furthermore, point source (Fig. 2d) and linear source (Fig. 2g) are placed in free space and further applied to demonstrate the effectiveness of the proposed concentrator in Fig. 2(d–f) and Fig. 2(g–i), respectively. The functionality of proposed devices as shown in Fig. 2(e–f) and Fig. 2(h,i) represents the concentration of the EM wave's energy in the central destined region. It can be clearly observed that the concentrator region is open to outer-world in all cases but functionality remains unchanged. Whereas, the location of point source is $(-0.3, 0.3)$ with the current density of $1A$ and on the other side, the height of linear source is 0.2 m at the origin point $(-0.4, 0)$ with the electric field intensity of $1V$.

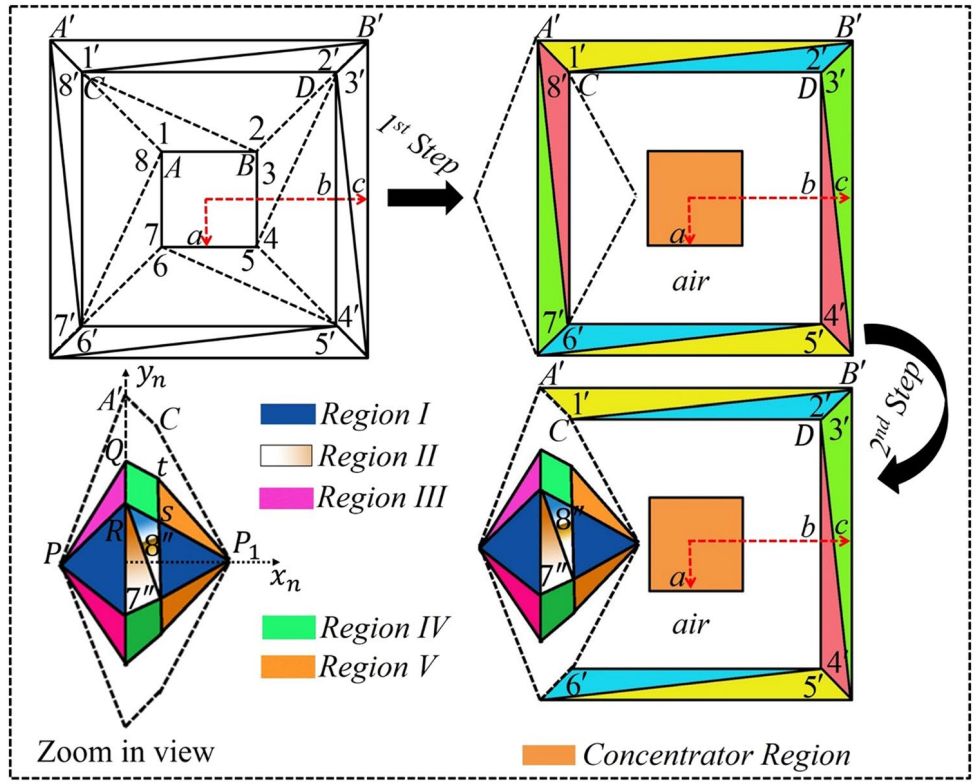

**Figure 1.** Scheme illustration of the proposed open concentration-coating device that concentrates EM waves while the concentrating region is not fully enclosed with the concentration-coating. The mapping of concentration-coating is defined as the dashed lines' regions in virtual space labeled by 1, 2, 3 …. 8 are transformed into a proposed concentrator with eight different regions (here labeled $1'$, $2'$, $3'$… $8'$). In easy words, the $\triangle ABC$ is folded into $\triangle A'B'C$ and so on. Meanwhile, the bigger square of radius $c$ is compressed into a smaller square region of radius $a$. It should be clearly noticed that the air gap exists between the coated material and concentrator region. In the second step, a multi-folded transformation method is applied to create a window. As an example, after applying the second transformation function, the regions $7'$ and $8'$ turns into $7''$ and $8''$, respectively, that is compressed and shifted towards the compressed region and that region is further coated with the folding regions.

A tradeoff is that the material parameters of the above example involves negative values of anisotropic and homogeneous materials that have two windows in-front of the concentrator region and these windows can be increased as per demand for remote controlling phenomenon. Since the designed concentrator can only store EM energy that raise a question that whether we are able to rotate and store EM energy simultaneously, for example, make a device that can be used for solar cells and for antenna applications as well. As an answer, we further extend the concept of open-coating concentrator into the open-coating rotator first and then combine these two devices to obtain the combinational effect.

The schematic diagram of the open-coating rotator is given in Fig. 3 with two different steps. At the first step, the $\triangle ABC$ of region 1 in virtual space $(x, y, z)$ is transformed into region $\triangle A'B'C'$ of label $1'$ in physical space $(x', y', z')$ and similarly, other regions in virtual space labeled by 2, 3 … 8 are accordingly transformed into colored regions 2′, 3′ ... 8′ to design the conventional rotator[20] made by homogenous materials. Apart from this particular step of transformation, the second step of this device is to achieve the isotropic materials of the open-coating device. In this regard, based on the specific choice[13] of $c = b^2/a$, the bigger dashed line circle of radius $c$ (the origin point is represented by black dot $O$) is compressed into region I of radius $a$ and that compressed region also contain some mapped regions. As a reference, we compressed the segments of 6′ and 7′ into the compressed region II and they became as 6″ and 7″, respectively. Therefore, the folding materials to cover the compressed region is obtained by folding the region $(c - b)$ into $(b - a)$.

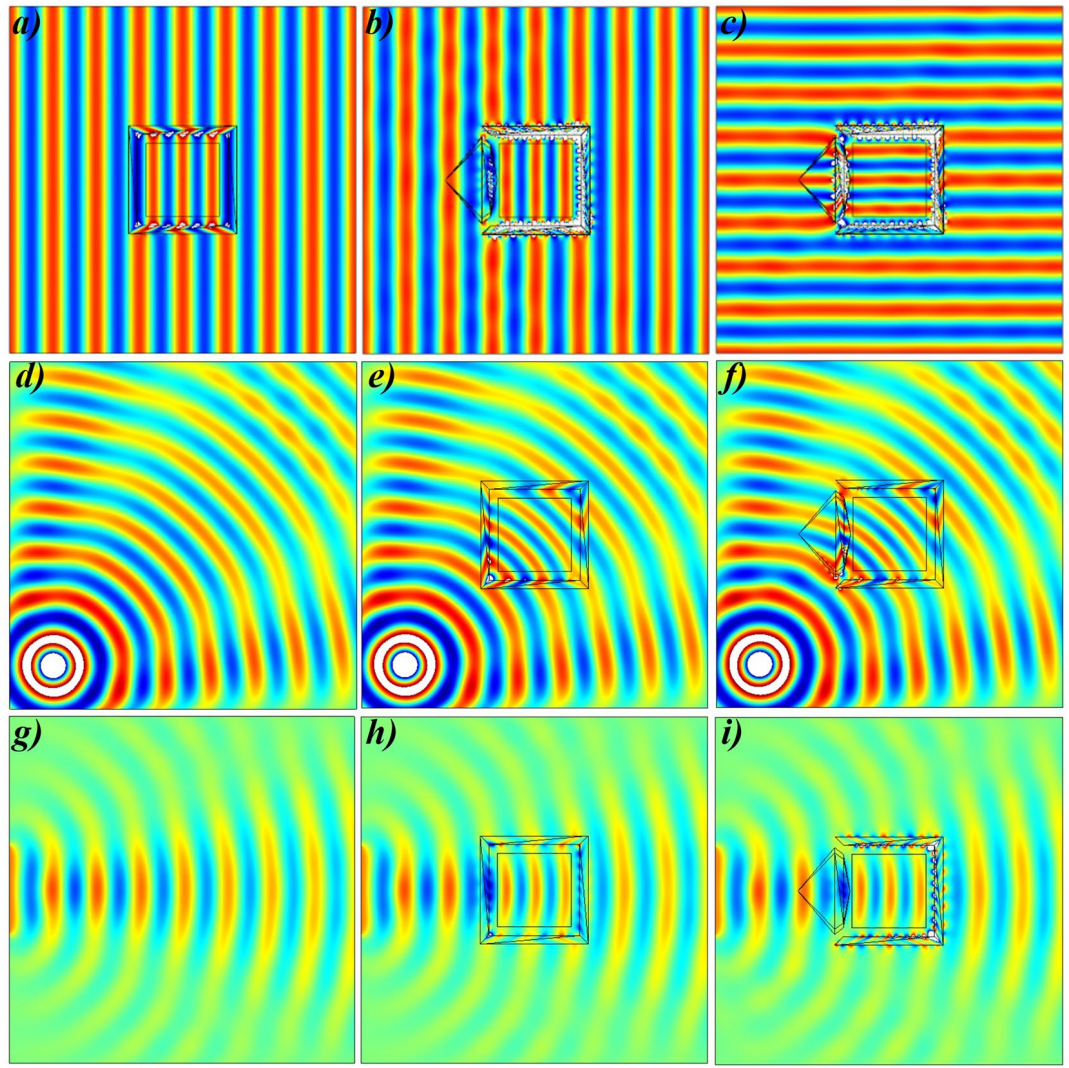

**Figure 2.** Simulation results of open-coating concentrator under different incident waves. (**a**) Electric field ($E_z$) distributions for the proposed concentrator under the illuminations of plane waves with incident angles of 0° and same incident waves are applied in (**b**) for the open-coating concentrator. (**c**) The incident angle of propagating waves are changed to 90° for (b). In all cases, the functionality of proposed concentrator is equivalent to previously designed TO-based concentrators[21]. (**d**) The EM waves of point source is concentrated in (**e**) closed concentrator and (**f**) open-concentrator. Similarly, emission of linear source in (**g**) is concentrated in (**h**) closed concentrator and (**i**) open-concentrator. In all cases, the results give the immediate validation of the proposed open-concentrator device.

Hereafter, the proposed designs are analyzed by using COMSOL Multiphysics in transverse electric (TE) mode at the frequency of $3\,GHz$. The geometric structure parameters for the both steps in SI units are as follows: the total width of the outer and inner squares are $0.35\,m$, $0.25\,m$, and the coordinates of $A = A' = (-0.175, 0.175)$, $B = B' = (0.175, 0.175)$, $C(-0.125, 0.125)$, $C'(0.125, 0.125)$, $O(-0.14, 0)$, $a = 0.04\,m$, $b = 0.05\,m$, and $c = 0.0625\,m$. The obtained material parameters for each region can be seen in method summary.

The simulation results are shown in Fig. 4(a–c). Figure 4a shows the rotator's field pattern in which the rotation region is fully encapsulated with the homogenous coating materials. Anyway, the functionality of our proposed rotator is equivalent to previously reported work[20]. Meanwhile, the Fig. 4b is the case of open-coating rotator in which the rotation region is open to the outer world. It can be clearly seen from Fig. 4(a,b) that the incident plane waves are impinging the devices with 0° angle and the devices rotate the EM waves with 90° angle in the destined region. In order to validate the bi-directional property of our proposed design, the incident wave is propagating along the 90° angle and as a result, the open-coating rotator's functionality is efficiently achieved, which can be seen in Fig. 4c.

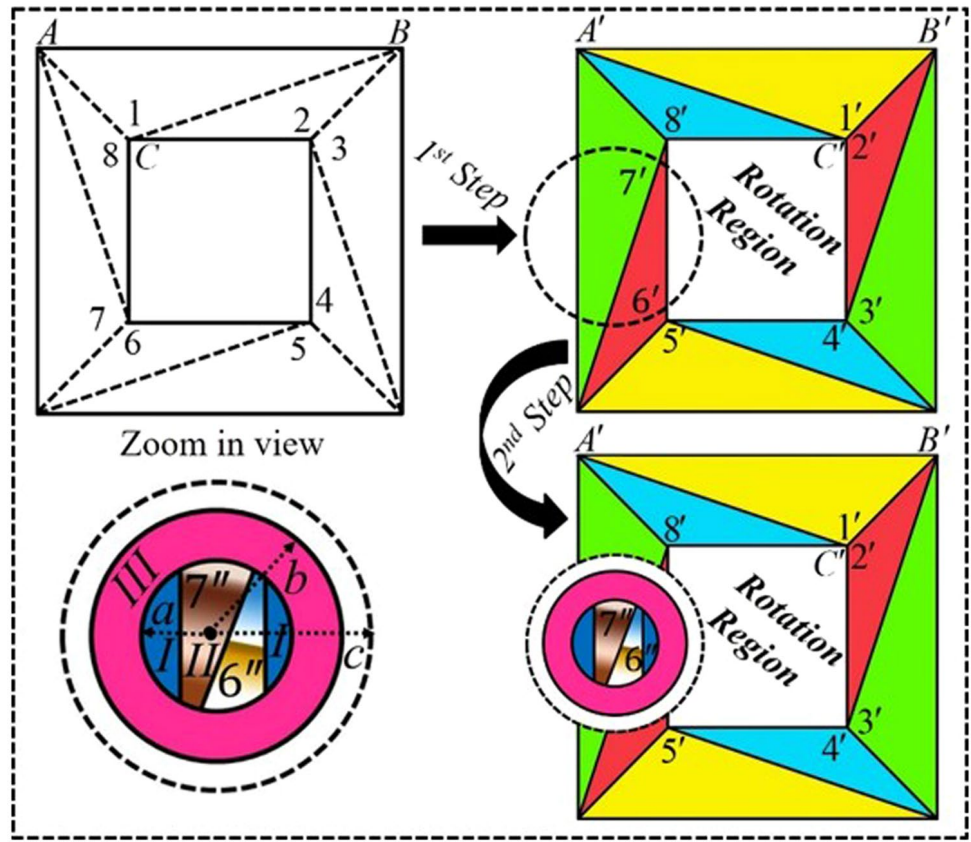

**Figure 3.** Scheme illustration of the proposed open rotation-coating device that rotates the EM field while the rotation region is not fully enclosed with the rotation-coating. A linear homogeneous coordinate transformation is used to design the rotator by constructing several triangle regions in virtual space. At the first step, the mapping of rotation-coating is defined as the dashed lines' regions in virtual space labeled by 1, 2, 3 …. 8 are transformed into a conventional rotator with eight different regions (here labeled 1′, 2′, 3′….8′). In easy words, the $\triangle ABC$ is mapped into $\triangle A'B'C'$ and so on. The material parameters of the proposed rotator designed from this method are homogeneous and easy to fabricate. In the second step, bigger dashed lines circle is compressed into region I and region II (contained mapped medium of 6′ and 7′ regions), and then isotropic folded material is used to coat the compressed region.

Compared with the open-coating concentrator as discussed previously in Fig. 1, in this case, the EM waves are rotated but not concentrated. By taking advantage from these two proposed designs, here, we plan to merge the first step of Fig. 1 into the first step of Fig. 3, which can be clarified in Fig. 5. In simple words, the outer coating layers are composed by rotators as the rotation region is free space, so in that free space the concentrator device is merged. The red lines are used to differentiate the concentrator and rotator boundaries. This bi-functional device is proposed to have the rotation and concentration of EM waves at the same time and we assume that device as a virtual space for the next step to make it like open-coating device. The procedure to make it open, again the second step of Fig. 3 is recalled. The geometric structure, constitutive parameters and working frequency of this particular device are same as described earlier for Figs 2, 4. For more details see method summary.

After that, simulations are performed and the results are shown in Fig. 6(a–c). Figure 6a validates the working performance of proposed bi-functional device that is rotating and concentrating the EM field at the same time in the destined region. In this case, the destined region is fully covered with the combination of proposed rotator (Fig. 3) and proposed concentrator (Fig. 1). Figure 6b is a case of open-coating device in which the destined region is open to the outer world. It can be observed in Fig. 6(a,b) that the incident plane waves are impinging the devices with 0° angle and the device rotate and concentrate the EM waves with 90° angle in the destined region. In Fig. 6c, the incident wave propagates along the 90° angle and the efficiency of proposed device remains unchanged.

It can be observed that there exist some negligible distortions in EM waves due to the resonance of the incident wave between the positive index material and negative index metamaterials to compensate each other.

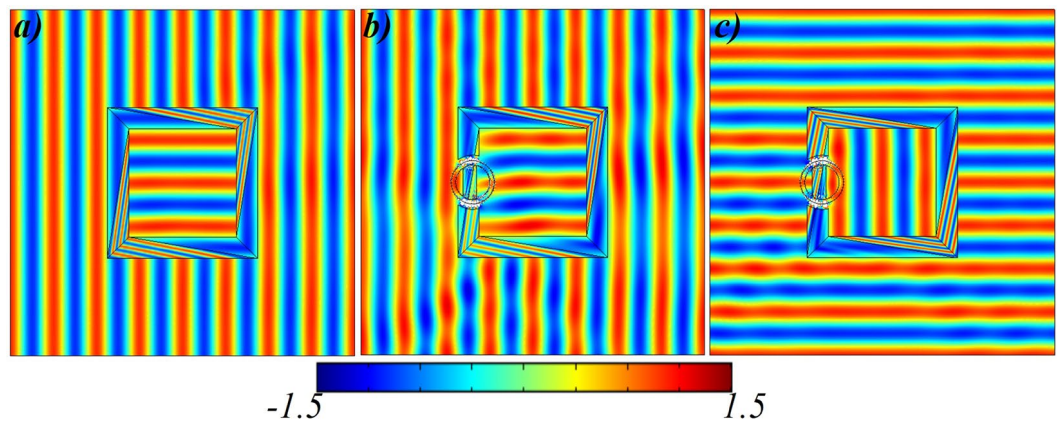

**Figure 4.** Simulation results of open-coating rotators. (**a**) Electric field ($E_z$) distributions for the proposed rotator under the illuminations of plane waves with incident angles of 0° and same incident waves are applied in (**b**) for the open-coating concentrator. (**c**) The incident angle of propagating waves are changed to 90° for (b). In all cases, the functionality of proposed rotator is equivalent to previously designed TO-based rotators[20].

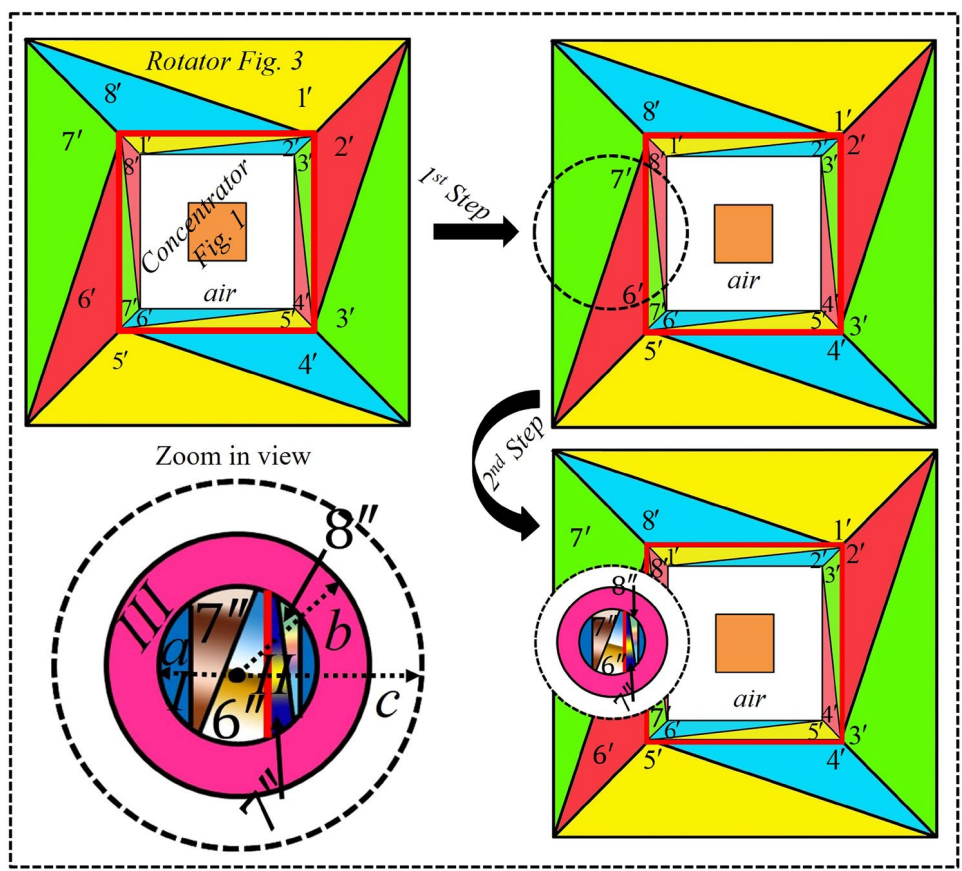

**Figure 5.** Scheme illustration of the proposed open-coating bi-functional device that first rotates and then concentrates the EM waves at the same time while the concentrator region is not fully enclosed with the coating materials. In the first step, the proposed bi-functional device is achieved by combining the first step of both Fig. 1 and Fig. 3, distinguished by the red lines boundaries. For second step, Fig. 3 (second step) has been recalled to make the proposed device open.

Theoretically, there should be no distortions according to the wave theory because of the existence of both folding space and resonance. However, for simulation case, it requires finest mesh to vanish the wave's distortions.

## Conclusion

Based on multi-folded TO, we proposed devices of homogeneous and anisotropic materials to manipulate the EM waves and further created air-gap between the coated materials in order to make destined region open to outer world. Meanwhile, the structure of open-coating devices are composed by mapped EM medium merged inside the compressed region that is further coated with negative index materials. The simulation results validated each case of different proposed structures and verified that the dielectric region can store and rotate the EM waves without any physical connectivity of coating materials. These features of proposed concept will be very helpful for future remote controlling applications in microwave and optical engineering.

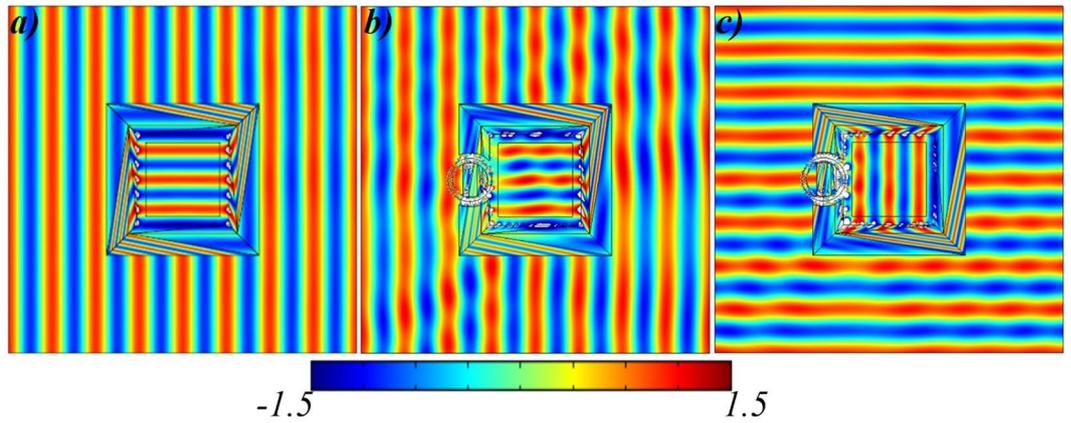

**Figure 6.** Simulation results of open-coating bi-functional device. (**a**) Electric field ($E_z$) distributions for the proposed bi-functional device under the illuminations of plane waves with incident angles of 0° and same incident waves are applied in (**b**) for the open-coating bi-functional device. (**c**) The incident angle of propagating waves are changed to 90° for (**b**).

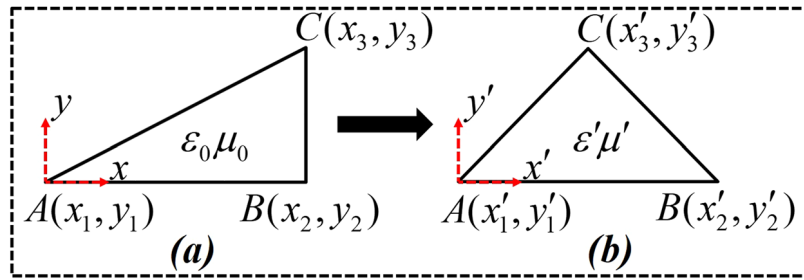

**Figure 7.** Representation of the linear homogeneous coordinate transformation in the virtual and real space. (**a**) The assumed triangular region of free space ($\varepsilon_0\mu_0$) in the Cartesian coordinate system is geometrically transformed into (**b**) another triangular shape.

## Methods

Before moving to the next steps, we consider the mapping shown in Fig. 7, which uniquely maps every point $(x, y)$ of Fig. 7a in a new region $(x', y')$ as shown in Fig. 7b. The transformation material of the new space $(x', y')$ can be related to its original space $(x, y)$ by:

$$\varepsilon' = \frac{J\varepsilon_0 J^T}{\det J}$$
$$\mu' = \frac{J\mu_0 J^T}{\det J} \tag{1}$$

where $J$ is the Jacobean tensor and for triangular shape only in 2D transformation, the Jacobean matrix can be found as:

$$J = \begin{bmatrix} x'_3 - x'_1 & x'_2 - x'_1 & 0 \\ y'_3 - y'_1 & y'_2 - y'_1 & 0 \\ 0 & 0 & 1 \end{bmatrix} \begin{bmatrix} x_3 - x_1 & x_2 - x_1 & 0 \\ y_3 - y_1 & y_2 - y_1 & 0 \\ 0 & 0 & 1 \end{bmatrix}^{-1} \tag{2}$$

In accordance to achieve the ultimate goal, in this paper, we focused on to achieve homogenous materials by using triangular shape regions. Thus, Eqs (1,2) will play an important role in designing the proposed devices.

We start from the first step of open-coating concentrator as mentioned in Fig. 1. Here, the bigger square of size $c = 0.25\ m$ is compressed into $a = 0.17\ m$ with the following transformation equations:

$$x' = \frac{a}{c}x$$
$$y' = \frac{a}{c}y$$
$$z' = z \qquad (3)$$

Thus, the constitutive parameters of that region will become as: $\varepsilon'_a = \mu'_a = diag[1, 1, (0.25/0.17)^2]$. In the same step, when $\Delta ABC$ of region 1 and $\Delta CDB$ of region 2 is folded into region $\Delta A'B'C$ named as $1'$ and $\Delta CDB'$ named as $2'$ respectively, the coordinates parameters of each point is known so this phenomenon is similar as mentioned in Fig. 7. So, by taking help from Eqs (1–2), the constitutive parameters will become as:

$$\varepsilon'_{1'} = \mu'_{1'} = \begin{bmatrix} -5.6212 & 1.68 & 0 \\ 1.68 & -0.68 & 0 \\ 0 & 0 & -0.68 \end{bmatrix}, \text{ and } \varepsilon'_{2'} = \mu'_{2'} = \begin{bmatrix} -5 & -2 & 0 \\ -2 & -1 & 0 \\ 0 & 0 & -1 \end{bmatrix}.$$

The same procedure is applied to find the material parameters of all other regions such as:

$$\varepsilon'_{3'} = \mu'_{3'} = \begin{bmatrix} -0.68 & -1.68 & 0 \\ -1.68 & -5.6212 & 0 \\ 0 & 0 & -0.68 \end{bmatrix}, \varepsilon'_{4'} = \mu'_{4'} = \begin{bmatrix} -1 & 2 & 0 \\ 2 & -5 & 0 \\ 0 & 0 & -1 \end{bmatrix},$$

$$\varepsilon'_{5'} = \mu'_{5'} = \begin{bmatrix} -5.6212 & 1.68 & 0 \\ 1.68 & -0.68 & 0 \\ 0 & 0 & -0.68 \end{bmatrix}, \varepsilon'_{6'} = \mu'_{6'} = \begin{bmatrix} -5 & -2 & 0 \\ -2 & -1 & 0 \\ 0 & 0 & -1 \end{bmatrix},$$

$$\varepsilon'_{7'} = \mu'_{7'} = \begin{bmatrix} -0.68 & -1.68 & 0 \\ -1.68 & -5.6212 & 0 \\ 0 & 0 & -0.68 \end{bmatrix}, \varepsilon'_{8'} = \mu'_{8'} = \begin{bmatrix} -1 & 2 & 0 \\ 2 & -5 & 0 \\ 0 & 0 & -1 \end{bmatrix}.$$

In the second step (Fig. 1), the bigger black dashed lines are compressed with the following transformation equations:

$$x' = x$$
$$y' = \kappa y$$
$$z' = z \qquad (4)$$

where $\kappa$ is the compression ratio of 0.7. So, the material properties of region I and region II will become as:

$$\varepsilon'_I = \mu'_I = diag[1/0.7, 0.7, 1/0.7],$$

$$\varepsilon''_{II7''} = \mu''_{II7''}\begin{bmatrix} -0.68/0.7 & -1.68 & 0 \\ -1.68 & -5.6212 \times 0.7 & 0 \\ 0 & 0 & -0.68/0.7 \end{bmatrix}, \varepsilon''_{II8''} = \mu''_{II8''}\begin{bmatrix} -1/0.7 & 2 & 0 \\ 2 & -5 \times 0.7 & 0 \\ 0 & 0 & -1/0.7 \end{bmatrix}.$$

After that folding transformation is applied and for each region as described in Fig. 1 (second step), the material parameters will become as:

$$\varepsilon'_{III} = \mu'_{III}\begin{bmatrix} -2 & -3.5294 & 0 \\ -3.5294 & -6.7284 & 0 \\ 0 & 0 & -2 \end{bmatrix}, \varepsilon'_{IV} = \mu'_{IV}\begin{bmatrix} -2 & 2.4 & 0 \\ 2.4 & -3.38 & 0 \\ 0 & 0 & -2 \end{bmatrix}, \text{ and } \varepsilon'_V = \mu'_V\begin{bmatrix} -2 & 16.80 & 0 \\ 16.80 & -141.62 & 0 \\ 0 & 0 & -2 \end{bmatrix}.$$

where as, the subscripts of $\varepsilon'$ and $\mu'$ indicate the material parameters of that concerned region only. It can be clearly observed from the constitutive parameters that the inhomogeneity of designed concentrator is completely removed, and only anisotropy is required to obtain the desired functionality.

For open-coating rotator (Fig. 3), in the first step, the conventional rotator is designed by simply mapping the region 1 as $\Delta ABC$ into region $\Delta A'B'C'$ of label $1'$ and similarly, other regions labeled as 2, 3 … 8 in virtual space are accordingly transformed into colored regions $2', 3'…8'$. Thus, by seeking help from Eqs (1,2), the constitutive parameters of the device will become as:

$$\varepsilon'_{1'} = \mu'_{1'}\begin{bmatrix} 26 & -5 & 0 \\ -5 & 1 & 0 \\ 0 & 0 & 1 \end{bmatrix}, \varepsilon'_{2'} = \mu'_{2'} = \begin{bmatrix} 1 & 7 & 0 \\ 7 & 50 & 0 \\ 0 & 0 & 1 \end{bmatrix}, \varepsilon'_{3'} = \mu'_{3'} = \begin{bmatrix} 1 & 5 & 0 \\ 5 & 26 & 0 \\ 0 & 0 & 1 \end{bmatrix}, \varepsilon'_{4'} = \mu'_{4'} = \begin{bmatrix} 50 & -7 & 0 \\ -7 & 1 & 0 \\ 0 & 0 & 1 \end{bmatrix},$$

$$\varepsilon'_{5'} = \mu'_{5'} = \begin{bmatrix} 26 & -5 & 0 \\ -5 & 1 & 0 \\ 0 & 0 & 1 \end{bmatrix}, \varepsilon'_{6'} = \mu'_{6'} = \begin{bmatrix} 1 & 7 & 0 \\ 7 & 50 & 0 \\ 0 & 0 & 1 \end{bmatrix}, \varepsilon'_{7'} = \mu'_{7'} = \begin{bmatrix} 1 & 5 & 0 \\ 5 & 26 & 0 \\ 0 & 0 & 1 \end{bmatrix}, \varepsilon'_{8'} = \mu'_{8'} = \begin{bmatrix} 50 & -7 & 0 \\ -7 & 1 & 0 \\ 0 & 0 & 1 \end{bmatrix}.$$

In the second step (Fig. 3), the bigger dashed lines circle is compressed into smaller blue colored circle with the following transformation equations:

$$x' = \frac{a}{c}x + x_0 - x_0\frac{a}{c}$$
$$y' = \frac{a}{c}y$$
$$z' = z \tag{5}$$

So, the material properties of region I and region II will become as:

$$\varepsilon'_I = \mu'_I = diag[1, 1, (c/a)^2], \quad \varepsilon''_{II7''} = \mu''_{II7''}\begin{bmatrix}1 & 5 & 0\\5 & 26 & 0\\0 & 0 & (\frac{c}{a})^2\end{bmatrix}, \quad \varepsilon''_{II6''} = \mu''_{II6''}\begin{bmatrix}1 & 7 & 0\\7 & 50 & 0\\0 & 0 & (\frac{c}{a})^2\end{bmatrix}.$$

Similarly the, material parameters of region III are obtained from ref.[13] that are: $\varepsilon'_{III} = \mu'_{III} = diag[-1, -1, -(b/r)^4]$ while $r = \sqrt{(x-x_0)^2 + y^2}$ and $x_0 = -0.14$.

For open-coating bi-functional device (Fig. 5), the constitutive parameters for both first and second step will remain similar as that of the obtained material parameters of Figs 1 and 3 except the folding regions. It should be noticed that the second step for this proposed design is achieved by recalling the Fig. 3 (second step), with the same geometric and material parameters except the value of $x_0 = -0.15$.

### Acknowledgements
This work was supported in part from the National Science Foundation of China under Grant Nos. 61631007, 61571117, 61501112, 61501117, 61522106, 61722106, 61701107, and 61701108, and 111 Project under Grant No.111-2-05. H.A. Madni acknowledges the support of the Higher Education Commission's Start-Up Research Grant Program, Pakistan under Grant No. 21-1742/SRGP/R&D/HEC/2017, and the support of the Postdoctoral Science Foundation of China at Southeast University, Nanjing, China, under Postdoctoral number 201557.

### Author Contributions
H.A. Madni designed the devices and carried out the simulations. K. Hussain, S. Liu, W.X. Jiang, A. Aziz, S. Iqbal, A. Mehboob and T.J. Cui analyzed the data and interpreted the results. H.A. Madni, S. Liu, W.X. Jiang, A. Mehboob and T.J. Cui draft the manuscript with the input from the others. T.J. Cui supervised the project.

### Additional Information
**Competing Interests:** The authors declare no competing interests.

**Publisher's note:** Springer Nature remains neutral with regard to jurisdictional claims in published maps and institutional affiliations.